\documentstyle[aps]{revtex}
\begin{document}
\draft
\author{Mahendra K. Verma and Jayant K. Bhattacharjee }
\address{Department of Physics, Indian Institute of Technology, \\ Kanpur
208016,
India}
\title{Direct Interaction Approximation of Magnetohydrodynamic Turbulence}
\date{\today}
\maketitle

\begin{abstract}
In this paper we apply Kraichnan's direct interaction approximation, which
is a one loop perturbation expansion, to magnetohydrodynamic turbulence. By
substituting the energy spectra both from kolmogorov-like MHD turbulence
phenomenology and a generalization of Dobrowolny et al.'s model we obtain
Kolmogorov's and Kraichnan's constant for MHD turbulence. We find that the
constants depend of the Alfv\'en ratio and normalized cross helicity; the
dependence has been studied here. We also demonstrate the inverse cascade of
magnetic energy for Kolmogorov-like models. Our results are in general
agreement with the earlier simulation results except for large normalized
cross helicity.
\end{abstract}

\pacs{PACS Numbers: 47.65.+a, 47.25.cg, 52.30}
\newpage

\section{INTRODUCTION}
Turbulence still remains primarily an unsolved problem. Among existing
statistical theories of fluid turbulence, Kolmogorov's phenomenology is the
most prominent one. Kolmogorov \cite{Kolm} hypothesized that the energy
spectrum $E(k)$ of fluid turbulence in the inertial range, i.e., for scales
between the energy feeding and dissipation range, is given by
\begin{equation}
E(k)=K_{Ko}\Pi ^{2/3}k^{-5/3},
\end{equation}
where $K_{Ko}$ is an universal constant called Kolmogorov's constant, $k$ is
the wavenumber, and $\Pi $ is the nonlinear cascade of energy, also equal to
the dissipation rate of the fluid. This power law has been confirmed by
experiments \cite{experiment} and numerical results \cite{simulation}.

Later on various theories were developed to understand fluid turbulence, the
primary ones being Direct Interaction Approximation (DIA) of Kraichnan \cite
{Kraich:59,Leslie}, Wyld's field-theoretic technique \cite{Wyld},
renormalization groups (RG) \cite{rg}, mode-coupling \cite{mode},
Eddy-Damped Quasi Normal Markovian (EDQNM) closure schemes \cite{closure}.
All these theories yield results consistent with the Kolmogorov's power law.
These calculations yield $K\approx 1.5$ in three dimensions (3D). In
two-dimensions (2D) Kraichnan \cite{Kraich:71}, Olla \cite{Olla}, and Nandy
and Bhattacharjee \cite{Nandy} show that $K_{Ko}\approx 6.4$ in region where
inverse cascade of energy occurs.

The DIA\ of Kraichnan \cite{Kraich:59,Leslie} is one of the fully consistent
analytical turbulence theory of fluid turbulence. It essentially involves
perturbation theory similar to that used in quantum field theory. In this
calculation Green's function and correlation function are calculated self
consistently to first nonvanishing order. Only triad interactions are
allowed to this order; for this reason it is called direct interaction
approximation (refer to Figure 1). The DIA of Kraichnan \cite{Kraich:59}
yielded
 $k^{-3/2}$ energy spectrum for fluid turbulence, which was found to
inconsistent with the experiments. However, later on it was shown that DIA
can also yield $k^{-5/3}$ energy spectrum \cite{Leslie}. The RG and mode
coupling theories are similar to DIA, and they have been shown to be yield
results close to those from DIA.

For magnetohydrodynamic (MHD) turbulence also there are several
phenomenologies. In MHD we have velocity field ${\bf u}$ and
magnetic field ${\bf B=B}_0+{\bf b}$, where ${\bf B}_0$ denotes the mean
magnetic field and ${\bf b}$ denotes the magnetic field fluctuations. In
this paper, in place of ${\bf u}$ and ${\bf b}$, we use Els\"asser variables
${\bf z}^{\pm }={\bf u\pm b/}\sqrt{4\pi \rho }$, where $\rho $ is the
density of the plasma. The wave speed of Alfv\'en waves due to the mean
magnetic field, called the Alfv\'en speed, is denoted by ${\bf C}_A$ and is
equal to ${\bf B}_0{\bf /}\sqrt{4\pi \rho }$. The variables ${\bf z}^{\pm }$
denote fluctuations having positive and negative velocity-magnetic field
correlations respectively. Throughout the paper we assume that $\rho $ is
constant, which implies that the fluid is incompressible. Marsch, Matthaeus
and Zhou, and Zhou and Matthaeus \cite{Kolm-like} proposed a phenomenology
similar to Kolmogorov's fluid turbulence phenomenology. We call this
Komogorov-like MHD turbulence phenomenology. In this phenomenology the
energy spectra $E^{\pm }(k)$ of the ${\bf z}^{\pm }$ fluctuations are
\begin{equation}
\label{Kolmlike}E^{\pm }(k)=K^{\pm }\left( \Pi ^{+}\right) ^{4/3}\left( \Pi
^{-}\right) ^{-2/3}k^{-5/3}
\end{equation}
where $\Pi ^{\pm }$ are the nonlinear energy cascades of ${\bf z}^{\pm }$
fluctuations, and $K^{\pm }$ are constants, whom we refer to as Kolmogorov's
constants for MHD turbulence. The spectral normalized cross helicity
$\sigma_{c}(k)$, defined as $(E^{+}(k)-E^{-}(k))/(E^{+}(k)+E^{-}(k))$, and
spectral
Alfv\'{e}n ratio $r_{A}(k)$, defined as $E^{u}/E^{b}$, play an important role
in MHD
turbulence. Here $E^{u}$ and $E^{b}$ are the kinetic and magnetic energies
respectively.  In  this paper $\sigma_{c}(k)$ and $r_{A}(k)$ are taken
to be independent of $k$, and we will usually
drop the term spectral while referring to these constants.
  According to the assumptions of the model,
this phenomenology is expected to be applicable when amplitudes of the
fluctuations are greater than the mean magnetic field or the magnetic field
of largest eddies, i.e., $z^{\pm }\gg B_0$. In this paper we theoretically
calculate the values of $K^{\pm }$. The quantity $E^R=\left( E^u-E^b\right) $%
, called the residual energy, plays an important role in MHD turbulence, as
will be seen later in this paper.

Kraichnan \cite{Kraich:mhd} argued that when the mean magnetic field or the
magnetic field of the largest eddies $C_A$ is much larger than the
fluctuations $z^{\pm }$, the magnetic energy spectrum $E^b(k)$ is given by
\begin{equation}
\label{Kraich}E^b(k)=A(\Pi ^bC_A)^{1/2}k^{-3/2},
\end{equation}
where $\Pi ^b$ is the magnetic energy cascade rate, and $A$ is Kraichnan's
constant. Dobrowolny et al. \cite{Dobro} generalized this model for ${\bf z}%
^{\pm }$ and found that
\begin{equation}
\label{Dobro}\Pi ^{+}=\Pi ^{-}=A^{-2}C_A^{-1}E^{+}(k)E^{-}(k)k^3,
\end{equation}
where $A$ is a constant. In this paper we also calculate the value of the
constant $A$ for a modified Dobrowolny et al.'s model.

Matthaeus and Zhou, and Zhou and Matthaeus \cite{Kolm-like} developed a
generalization that contains both Kolmogorov-like and Dobrowolny et al.'s
models as limiting cases. In their model the dissipation rates are given by
\begin{equation}
\label{ZM}\Pi ^{\pm }=\frac{A^{-2}B_0^{-1}E^{+}(k)E^{-}(k)k^3}{C_A+\sqrt{%
kE^{\pm }(k)}}.
\end{equation}
In this paper we do not discuss this model because of its complexity.

Some of the analytical theories of fluid turbulence have been applied to MHD
turbulence. For example, Pouquet et al. and Grappin et al. \cite{Grappin}
applied the EDQNM closure schemes, and Fournier et al. and Camargo and
Tasso \cite{mhdrg} have applied the RG technique to MHD turbulence. Verma
and Bhattacharjee \cite{Verma:dia} have applied DIA technique to MHD
turbulence when $E^{-}(k)=E^{+}(k)$ and obtained the Kolmogorov's constant $K
$. In this paper we have generalized their formalism for cases when $%
E^{-}(k)\neq E^{+}(k)$ and obtained $K^{\pm }$. In this paper
we have also theoretically
obtained Kraichnan's constant $A$ for a modified Dobrowolny et al.'s model.

The solar wind is an ideal natural experiment where some of the predictions
of MHD turbulence theories could be tested. The solar wind observations by
the spacecraft show that energy spectra of the solar wind is closer to $%
k^{-5/3}$ rather than to $k^{-3/2}$ \cite{observation}. This is in spite of
the fact that the mean magnetic field in the solar wind (approximately 5
nanotesla) is  approximately 2 to 10 times larger than the large-scale
fluctuations. However, it should be noted that the exponents $5/3$ and $3/2$
are difficult to distinguish. Recently Verma et al. \cite{Verma:sim} have
performed direct numerical simulations of MHD turbulence in which they could
not conclude whether the spectral index was 5/3 or 3/2. However, in their
simulation the nonlinear energy cascade rates $\Pi ^{\pm }$ appear to follow
Eq.~(\ref{Kolmlike}) rather than Eq.~(\ref{Dobro}). Note that the EDQNM
calculations of Grappin et al. \cite{Grappin} show that the spectral index
varies between 0 and 3 depending on normalized cross helicity.

The outline of the paper is as follows: In section 2 of this paper we derive
the equations for the correlations functions and flux functions that are
used in subsequent sections. In section 3 we show that the energy spectra of
Eq.~(\ref{Kolmlike}) are consistent solutions of MHD equations and find the
constants $K^{\pm }$. We find the constants for various $E^{-}(k)/E^{+}(k)$
and $E^u(k)/E^b(k)$ ratios. In section 4 we construct a generalization of
Dobrowolny et al.'s model, and then calculate the
Kraichnan's constant that model. Section 5
contains discussion and conclusions. Other than a brief remarks in sections
4, we do not attempt to address the important question under what ratio of $%
z^{\pm }/B_0$ the transitions from the energy spectra $k^{-5/3}$ to $k^{-3/2}
$ take place; the answer to this question require further analysis which is
beyond the scope of this paper.

\section{MHD TURBULENCE}

The MHD equations in terms of ${\bf z}^{\pm }$ are \cite{Kraich:mhd}
\begin{equation}
\frac \partial {\partial t}{\bf z}^{\pm }\mp {\bf C}_A\cdot \nabla {\bf z}%
^{\pm }+{\bf z}^{\mp }\cdot \nabla {\bf z}^{\pm }=-\frac 1\rho \nabla
p^{tot}+\nu _{+}\nabla ^2{\bf z}^{\pm }+\nu _{-}\nabla ^2{\bf z}^{\mp }+{\bf %
f}^{\pm },
\end{equation}
\begin{equation}
\nu _{\pm }=\frac{\nu \pm c^2/(4\pi \sigma )}2
\end{equation}
where ${\bf C}_A={\bf B}_0/\sqrt{4\pi \rho }$, $p^{tot}$ is the total
pressure, i.e., thermal plus magnetic pressure, ${\bf f}^{\pm }$ are
forcing, $\nu $ is the kinematic viscosity, $c$ is the speed of light, and $%
\sigma $ is conductivity. The corresponding equation in Fourier space is
\begin{equation}
\label{mhdeqn0}
\begin{array}{c}
\left(
\frac d{dt}+\nu _{+}k^2\right) z_i^{\pm }({\bf k},t)+\nu _{-}k^2z_i^{\mp }(%
{\bf k},t)\mp i({\bf C}_A\cdot {\bf k)}z_i^{\pm }({\bf k},t)= \\ f_i^{\pm }(%
{\bf k})-ik_ip^{tot}({\bf k},t)-\int d{\bf p}p_jz_j^{\mp }({\bf p}%
,t)z_i^{\pm }({\bf k-p},t)
\end{array}
\end{equation}
We can eliminate pressure $p^{tot}$ using the incompressibility condition,
which is
$$
k_iz_i^{\pm }({\bf k},t)=0.
$$
After the elimination of pressure the incompressible equation is
\begin{equation}
\label{mhdeqn}
\begin{array}{c}
\left(
\frac d{dt}+\nu _{+}k^2\right) z_i^{\pm }({\bf k},t)+\nu _{-}k^2z_i^{\mp }(%
{\bf k},t)\mp i({\bf C}_A\cdot {\bf k)}z_i^{\pm }({\bf k},t)= \\ f_i^{\pm }(%
{\bf k})-\epsilon M_{ijm}({\bf k})\int d{\bf p}z_j^{\mp }({\bf p},t)z_m^{\pm
}({\bf k-p},t)
\end{array}
\end{equation}
where
\begin{equation}
M_{ijm}({\bf k})=k_jP_{im}({\bf k});\ P_{im}({\bf k})=\delta _{im}-\frac{%
k_ik_m}{k^2}.
\end{equation}
We have introduced an expansion parameter $\epsilon $ which is set to one at
the end. We assume that the forcing $f^{+}$ and $f^{-}$ are independent,
i.e., $\partial f^{s_1}/\partial f^{s_2}=\delta _{s_1s_2}$, where $%
s_1,s_2=\pm $. Note that when ${\bf z}^{+}={\bf z}^{-}={\bf u}$, we recover
the Navier-Stokes equation. The quantity $M_{ijm}({\bf k})$ for MHD is
different
from that in fluid turbulence because of asymmetry introduced by ${\bf z}^{+}
$ and ${\bf z}^{-}$.

Now we define $z-z$ correlation tensors $S_{ij}$ as follows:
\begin{equation}
\left\langle z_i^{+}({\bf k},t)z_j^{+}({\bf k}^{\prime },t^{\prime
})\right\rangle =S_{ij}^{++}({\bf k,}t,t^{\prime }{\bf )}\delta ({\bf k+k}%
^{\prime }),
\end{equation}
\begin{equation}
\left\langle z_i^{-}({\bf k},t)z_j^{-}({\bf k}^{\prime },t^{\prime
})\right\rangle =S_{ij}^{--}({\bf k,}t,t^{\prime }{\bf )}\delta ({\bf k+k}%
^{\prime }),
\end{equation}
\begin{equation}
\left\langle z_i^{+}({\bf k},t)z_j^{-}({\bf k}^{\prime },t^{\prime
})\right\rangle =S_{ij}^{+-}({\bf k,}t,t^{\prime }{\bf )}\delta ({\bf k+k}%
^{\prime }).
\end{equation}
Here $\left\langle x\right\rangle $ denotes an ensemble average of $x$. We
assume that the turbulence is stationary, which implies that the
correlations are independent of time: $S_{ij}^{s_1s_2}({\bf k,}t,t{\bf )=}$ $%
S_{ij}^{s_1s_2}({\bf k)}$.  For fluid turbulence
Orszag \cite{closure} derived correlation functions and their properties.
Our following discussion is similar to that of Orszag.

{}From Eq.~(\ref{mhdeqn}) we can obtain equations for correlation tensors
that
are
\begin{equation}
\label{corrpm}
\begin{array}{c}
\left(
\frac d{dt}+2\nu _{+}k^2\right) S_{ij}^{\pm \pm }( {\bf k},t,t)+2\nu
_{-}k^2S_{ij}^{+-}({\bf k},t,t)-2i({\bf C}_A\cdot {\bf k)}S_{ij}^{\pm \pm }(%
{\bf k},t,t)= \\ 2F_{ij}^{\pm }({\bf k)}-\epsilon M_{ilm}({\bf k})\int d{\bf %
p}T_{jlm}^{\pm }({\bf -k,p)}-\epsilon M_{jlm}(-{\bf k})\int d{\bf p}%
T_{ilm}^{\pm }({\bf k,p)}
\end{array}
\end{equation}
\begin{equation}
\label{corrR}
\begin{array}{c}
\left(
\frac d{dt}+2\nu _{+}k^2\right) S_{ij}^{+-}( {\bf k},t,t)+\nu _{-}k^2\left(
S_{ij}^{++}({\bf k},t,t)+S_{ij}^{--}({\bf k},t,t)\right) \\ -2i(
{\bf C}_A\cdot {\bf k)}S_{ij}^{+-}({\bf k},t,t)=2F_{ij}^R({\bf k)} \\
-\epsilon M_{ilm}({\bf k})\int d{\bf p}T_{jlm}^{R1}(-{\bf k,p)}-\epsilon
M_{jlm}(-{\bf k})\int d{\bf p}T_{ilm}^{R2}({\bf k,p)}
\end{array}
\end{equation}
where
\begin{equation}
\left\langle z_j^{\pm }({\bf k},t)z_l^{\mp }({\bf p},t)z_m^{\pm }({\bf q}%
,t)\right\rangle =T_{jlm}^{\pm }({\bf k,p)}\delta ({\bf k+p+q)}
\end{equation}
\begin{equation}
\left\langle z_j^{-}({\bf k},t)z_l^{-}({\bf p},t)z_m^{+}({\bf q}%
,t)\right\rangle =T_{jlm}^{R1}({\bf k,p)}\delta ({\bf k+p+q)}
\end{equation}
\begin{equation}
\left\langle z_j^{+}({\bf k},t)z_l^{+}({\bf p},t)z_m^{-}({\bf q}%
,t)\right\rangle =T_{jlm}^{R2}({\bf k,p)}\delta ({\bf k+p+q)}
\end{equation}
\begin{equation}
\left\langle f_i^{\pm }({\bf k})z_j^{\pm }({\bf k}^{\prime }{\bf )}%
\right\rangle =F_{jm}^{\pm }({\bf k)}\delta ({\bf k+k}^{\prime })
\end{equation}
\begin{equation}
\frac 12\left\langle f_i^{+}({\bf k})z_j^{-}({\bf k}^{\prime }{\bf )+}%
f_i^{-}({\bf k})z_j^{+}({\bf k}^{\prime }{\bf )}\right\rangle =F_{jm}^R({\bf %
k)}\delta ({\bf k+k}^{\prime })
\end{equation}

For isotropic turbulence
\begin{equation}
S_{ij}^{s_1s_2}({\bf k)=}P_{ij}({\bf k})C^{s_1s_2}(k).
\end{equation}
In presence of a mean magnetic field, the correlation tensors are expected to
be
anisotropic. In the solar wind the fluctuations are anisotropic; the
amplitudes of the fluctuations along the Parker field is one-third that of
those perpendicular to the Parker field \cite{Roberts}. However, the
analysis of anisotropic turbulence is quite complicated, and in this paper
we assume isotropy for all the correlations.

We also define one-dimensional energy spectra as
\begin{equation}
\label{Ek3d}E^{\pm }(k)=\frac{C^{\pm \pm }(k)}{4\pi k^2};E^R(k)=\frac{%
C^{+-}(k)}{4\pi k^2}:in3D
\end{equation}
\begin{equation}
\label{Ek2d}E^{\pm }(k)=\frac{C^{\pm \pm }(k)}{2\pi k};E^R(k)=\frac{C^{+-}(k)%
}{2\pi k}:in2D
\end{equation}
For isotropic distribution in $k$ space, after substitution of $i=j$ in Eqs.
(\ref{corrpm},\ref{corrR}) and taking the trace, we obtain
\begin{equation}
\label{Tpm}\left( \frac d{dt}+2\nu _{+}k^2\right) E^{\pm }({\bf k},t,t)+2\nu
_{-}k^2E^R({\bf k},t,t)=T^{\pm }(k,t)
\end{equation}
\begin{equation}
\label{TR}\left( \frac d{dt}+2\nu _{+}k^2\right) E^R({\bf k},t,t)+\nu
_{-}k^2\left( E^{+}({\bf k},t,t)+E^{-}({\bf k},t,t)\right) =T^R(k,t)
\end{equation}
where
\begin{equation}
\label{Tpm2}T^{\pm }(k,t)=-4\pi k^2M_{ijm}({\bf k)}Im\int d{\bf p}%
T_{ijm}^{\pm }({\bf k,p)}
\end{equation}
\begin{equation}
\label{TR2}T^R(k,t)=-4\pi k^2M_{ijm}({\bf k)}Im\int d{\bf p}\left( \frac{%
T_{ijm}^{R2}({\bf k,p)-}T_{ijm}^{R1}({\bf k,p)}}2\right) .
\end{equation}
Note that the term $({\bf C}_A\cdot {\bf k)}E^{\pm }(k)=$ $C_Ak_sE^{\pm }(k)$%
, where $k_s$ is the wavevector along the mean magnetic field, is zero for
isotropic turbulence because it is an odd function of $k_s$.

When $\nu _{+}=\nu _{-}=0$, one can easily check from the  above equations
that the total energy $E^{\pm }=\int E^{\pm }(k)dk$ is conserved. However,
total residual energy $E^R=\int E^R(k)dk$ is not conserved. In terms of $T$%
's these statements of energy conservation are
\begin{equation}
\int_0^\infty T^{\pm }(k,t)dk=0
\end{equation}
but
\begin{equation}
\int_0^\infty T^R(k,t)dk\neq 0.
\end{equation}

We can also derive a theorem of ``detailed conservation of energy $E^{\pm
}(k)$'', i.e.,
\begin{equation}
s^{\pm }({\bf k,p,q})+s^{\pm }({\bf k,q,p})+s^{\pm }({\bf p,q,k})+s^{\pm }(%
{\bf p,k,q})+s^{\pm }({\bf q,k,p})+s^{\pm }({\bf q,p,k})=0,
\label{detail}
\end{equation}
where
\begin{equation}
s^{\pm }({\bf k,p,q})\delta ({\bf k+p+q)=-}Im\left\langle \left( {\bf k\cdot
z}^{\mp }({\bf p)}\right) \left( {\bf z}^{\pm }({\bf k)\cdot z}^{\pm }({\bf %
q)}\right) \right\rangle
\end{equation}
The diagrammatic representation of the terms $S^{+}(\cdot ,\cdot ,\cdot )$
is shown in Figure 1. The solid lines represent $z^{+}$ and the wavy line
represent $z^{-}$. For $S^{-}(\cdot ,\cdot ,\cdot )$ we need to change the
solid lines to wavy lines and vice versa. The physical interpretation of
these diagrams is that the energy from any $z^{+}({\bf k)}$ mode gets
transferred to modes $z^{+}({\bf p)}$ and $z^{-}({\bf q)}$ (${\bf k+q=k}$),
and so does energy from $z^{-}({\bf k)}$. However, the quantities $E^{+}(%
{\bf k)+}$ $E^{+}({\bf p)+}$ $E^{+}({\bf q)}$ and $E^{-}({\bf k)+}$ $E^{-}(%
{\bf p)+}$ $E^{-}({\bf q)}$ are separately conserved.

In presence of viscosity there is dissipation in the plasma. The dissipation
rates of $z^{\pm }$ fluctuations are given by
\begin{equation}
\begin{array}{c}
\varepsilon ^{\pm }=-
\frac d{dt}\int_0^\infty E^{\pm }(k,t)dk \\ =2\nu _{+}\int_0^\infty
k^2E^{\pm }(k,t)dk+2\nu _{-}\int_0^\infty k^2E^R(k,t)dk.
\end{array}
\end{equation}
Integrating both sides of Eqs. (\ref{Tpm},\ref{TR}) over a sphere of radius $%
K$ in $k$ space, we obtain the energy cascade rates $\Pi ^{\pm ,R}(K,t)$
that is

\begin{equation}
\label{fluxpm}
\begin{array}{c}
\Pi ^{\pm }(K,t)=-\int_0^KT^{\pm }(k,t)dk=\int_K^\infty T^{\pm }(k,t)dk \\
=-\frac d{dt}\int_0^KE^{\pm }(k,t)dk-2\nu _{+}\int_0^\infty k^2E^{\pm
}(k,t)dk-2\nu _{-}\int_0^\infty k^2E^{\pm }(k,t)dk.
\end{array}
\end{equation}
\begin{equation}
\label{fluxR}
\begin{array}{c}
\Pi ^R(K,t)=-\int_0^KT^R(k,t)dk \\
=-\frac d{dt}\int_0^KE^R(k,t)dk-2\nu _{+}\int_0^\infty k^2E^{\pm
}(k,t)dk-2\nu _{+}\int_0^\infty k^2E^{\pm }(k,t)dk.
\end{array}
\end{equation}

The above equations will be used in the subsequent sections. In Eqs. (\ref
{Tpm},\ref{TR}) we find that the $n$th-order moments involves $(n+1)$%
th-order moments along with $n$th and lower order moments. Hence one cannot
close these equations. This is the famous ``closure problem'' \cite{closure}%
. \ To circumvent this difficulty, various techniques have been devised,
primary ones being renormalization groups \cite{rg,mhdrg}, DIA \cite
{Kraich:59,Leslie}, mode-coupling scheme \cite{mode}, EDQNM closure schemes
\cite{closure,Grappin} etc. In this paper we only discuss DIA\ of MHD
turbulence.

\section{DIA\ OF MHD WITH $C_A=0$}

In this section we assume that the mean magnetic field or the magnetic field
due to the large eddies is small as compared to the fluctuations, i.e., $%
C_A\ll z^{\pm }.$ Here we set $C_A=0$. In the next section we will analyse
for the case when $C_A\gg z^{\pm }.$ Firstly, we calculate the Green's
function in the spirit of Kraichnan \cite{Kraich:59,Leslie}.

\subsection{Computation of Green's Function}

It is convenient to derive Green's function in Fourier space $\hat k=({\bf k}%
,\omega ).$ Following Leslie \cite{Leslie} we define Green's function $%
G_{ij}^{s_1s_2}(\hat k)$ as
\begin{equation}
G_{ij}^{s_1s_2}(\hat k)=\frac{\delta z_i^{s_1}(\hat k)}{\delta f_j^{s_2}}.
\end{equation}
In Fourier space the equation of Green's function is
\begin{equation}
\label{Green}
\begin{array}{c}
\left[
\begin{array}{cc}
-i\omega +\nu _{+}k^2 & \nu _{-}k^2 \\
\nu _{-}k^2 & -i\omega +\nu _{+}k^2
\end{array}
\right] \left[
\begin{array}{cc}
G_{ij}^{++}(\hat k) & G_{ij}^{+-}(
\hat k) \\ G_{ij}^{-+}(\hat k) & G_{ij}^{--}(\hat k)
\end{array}
\right] =P_{ij}(
{\bf k)}-\epsilon M_{ilm}({\bf k)\times } \\
\displaystyle \int
_{\hat p+\hat q=\hat k}d\hat p\left[
\begin{array}{cc}
z_l^{-}(\hat p) & z_m^{+}(
\hat q) \\ z_l^{-}(\hat p) & z_m^{+}(\hat q)
\end{array}
\right] \left[
\begin{array}{cc}
G_{mj}^{++}(\hat q) & G_{mj}^{+-}(
\hat p) \\ G_{lj}^{-+}(\hat q) & G_{lj}^{--}(\hat p)
\end{array}
\right]
\end{array}
\end{equation}
where $\hat p=({\bf p,}\omega ^{\prime })$ and $\hat q=({\bf k-p,\omega -}%
\omega ^{\prime })$. We denote
\begin{equation}
\left[
\begin{array}{cc}
-i\omega +\nu _{+}k^2 & \nu _{-}k^2 \\
\nu _{-}k^2 & -i\omega +\nu _{+}k^2
\end{array}
\right] ^{-1}=\hat G^0({\bf k},\omega )
\end{equation}
\begin{equation}
\hat G_{in}({\bf k},\omega )=P_{in}({\bf k})\hat G(\hat k)
\end{equation}

Now we solve the Eqs. (\ref{mhdeqn}) and (\ref{Green}) perturbatively. We
expand ${\bf z}^{\pm }$ and Green's function $\hat G$ in series:
\begin{equation}
z_i^{\pm }(\hat k)=z_i^{\pm (0)}(\hat k)+\epsilon z_i^{\pm (1)}(\hat k%
)+O(\epsilon ^2)
\end{equation}
\begin{equation}
G_{ij}^{s_1s_2}(\hat k)=G_{ij}^{s_1s_2(0)}(\hat k)+\epsilon
G_{ij}^{s_1s_2(1)}(\hat k)+O(\epsilon ^2)
\end{equation}
In DIA we keep terms only up to first nonvanishing order \cite{Leslie}. By
substituting these terms in Eqs. (\ref{mhdeqn}) and (\ref{Green}), and
equating the terms with equal powers of $\epsilon $, we obtain
\begin{equation}
\label{z0}\left[
\begin{array}{c}
z_i^{+(0)}(
\hat k) \\ z_i^{-(0)}(\hat k)
\end{array}
\right] =\hat G^0(\hat k)\left[
\begin{array}{c}
f_i^{+} \\
f_i^{-}
\end{array}
\right]
\end{equation}
\begin{equation}
\label{z1}\left[
\begin{array}{c}
z_i^{+(1)}(
\hat k) \\ z_i^{-(1)}(\hat k)
\end{array}
\right] =-i\hat G^0(\hat k)M_{ijk}({\bf k)\times }\int_{\hat p+\hat q=\hat k%
}d\hat p\left[
\begin{array}{c}
z_j^{-(0)}(
\hat p)z_m^{+(0)}(\hat q) \\ z_j^{+(0)}(\hat p)z_m^{-(0)}(\hat q)
\end{array}
\right]
\end{equation}
\begin{equation}
\label{G1}\hat G^{(1)}(\hat k)=-\frac i2\hat G_{il}^0(\hat k)M_{ljm}({\bf k)}%
\displaystyle \int
_{\hat p+\hat q=\hat k}d\hat p\left[
\begin{array}{cc}
z_j^{-(0)}(\hat p) & z_m^{+(0)}(
\hat q) \\ z_j^{-(0)}(\hat p) & z_m^{+(0)}(\hat q)
\end{array}
\right] \left[
\begin{array}{cc}
G_{mi}^{++(0)}(\hat q) & G_{mi}^{+-(0)}(
\hat q) \\ G_{ji}^{-+(0)}(\hat p) & G_{ji}^{--(0)}(\hat p)
\end{array}
\right]
\end{equation}
The factor 1/2 in the equation appears because of trace of $P_{ij}(k)$.
In DIA ${\bf f}$ is assumed to have a gaussian distribution. Therefore,
according to Eq. (\ref{z0}), ${\bf z}^{\pm (0)}$ also have a Gaussian
distribution. Green's function $\hat G$ to leading order is (refer to Eq.
\ref{Green})
\begin{equation}
\begin{array}{c}
\hat G(\hat k)=\hat G^{(0)}(\hat k)-\frac i2\epsilon ^2\hat G_{il}^{(0)}(%
\hat k)M_{ljm}({\bf k)}%
\displaystyle \int
_{\hat p+\hat q=\hat k}d\hat p \\ \left\langle \left[
\begin{array}{cc}
z_j^{-(0)}(\hat p) & z_m^{+(0)}(
\hat q) \\ z_j^{-(0)}(\hat p) & z_m^{+(0)}(\hat q)
\end{array}
\right] \left[
\begin{array}{cc}
G_{mi}^{++(1)}(\hat q) & G_{mi}^{+-(1)}(
\hat q) \\ G_{ji}^{-+(1)}(\hat p) & G_{ji}^{--(1)}(\hat p)
\end{array}
\right] \right\rangle
\end{array}
\end{equation}
In the above equation we have used the fact that $\hat G^{(0)}({\bf k}%
,\omega )$ is statistically sharp \cite{Leslie}, and $\left\langle z_i^{\pm
(0)}\right\rangle =0$, $\left\langle ZG\right\rangle =\left\langle
Z^{(0)}G^{(1)}\right\rangle +O(\epsilon ^2)$. The substitution of $G^1$ of
Eq. (\ref{G1}) in the above equation yields
\begin{eqnarray}
\hat G(\hat k) & = & \hat G^0(\hat k)  -  \frac i2\hat G_{il}^0(\hat
k)M_{ljm}({\bf k)}%
\displaystyle \int
_{\hat p+\hat q=\hat k}d\hat p <\left[
\begin{array}{cc}
z_j^{-(0)}(\hat p) & z_m^{+(0)}(
\hat q) \\ z_j^{-(0)}(\hat p) & z_m^{-(0)}(\hat q)
\end{array}
\right] \times  \nonumber \\
 & & \left[
\begin{array}{cc}
M_{sab}(\hat q)G_{ms}^{++(0)}(\hat q)%
\displaystyle \int
_{\hat r+\hat s=\hat q}d\hat r & M_{sab}(
\hat q)G_{ms}^{+-(0)}(\hat q)%
\displaystyle \int
_{\hat r+\hat s=\hat q}d\hat r \\ M_{sab}(\hat p)G_{js}^{-+(0)}(\hat p)%
\displaystyle \int
_{\hat r+\hat s=\hat p}d\hat r & M_{sab}(
\hat p)G_{js}^{--(0)}(\hat p)%
\displaystyle \int
_{\hat r+\hat s=\hat p}d\hat r
\end{array} \right] \times  \nonumber \\
& & \left[
\begin{array}{cc}
z_j^{-(0)}(\hat r)P_{bn}(\hat s) & z_m^{+(0)}(
\hat s)P_{an}(\hat r) \\ z_j^{-(0)}(\hat r)P_{bn}(\hat s) & z_m^{-(0)}(\hat s%
)P_{an}(\hat r)
\end{array}
\right] \left[
\begin{array}{cc}
G^{++(0)}(\hat s) & G_{mi}^{+-(0)}(
\hat s) \\ G_{ji}^{-+(0)}(\hat r) & G_{ji}^{--(0)}(\hat r)
\end{array}
\right] >
\label{Gfull}
\end{eqnarray}
Following Kraichnan's DIA \cite{Kraich:59,Leslie} we substitute $\epsilon =1$
and replace $z^{(0)}$ by $z$ and $G^{(0)}$ in 2nd and 3rd line by $G$.
Comparing the resulting equation with
\begin{equation}
\label{Gdef}\hat G(\hat k)=\hat G^0(\hat k)-\hat G^0(\hat k)\hat \Sigma (%
\hat k)\hat G^0(\hat k)
\end{equation}
where $\Sigma $ is self-energy, we obtain
\begin{equation}
\label{sigma}\hat \Sigma (\hat k)=\left[
\begin{array}{cc}
\Sigma ^{++}(\hat k) & \Sigma ^{+-}(
\hat k) \\ \Sigma ^{-+}(\hat k) & \Sigma ^{--}(\hat k)
\end{array}
\right] =\frac 12\left[
\begin{array}{cc}
\eta ^{+}(\hat k) & \eta ^{-}(
\hat k) \\ \eta ^{+}(\hat k) & \eta ^{-}(\hat k)
\end{array}
\right] ,
\end{equation}
\begin{equation}
\label{sigp}
\begin{array}{c}
\eta ^{+}(
\hat k)=\frac 12k^2\int_{\hat p+\hat q=\hat k}d\hat p\times  \\ \left[
b_1(k,p,q)\left( G^{-+}(\hat p)+G^{--}(\hat p)\right) C^{+-}(\hat q%
)+b_2(k,p,q)\left( G^{++}(\hat p)+G^{+-}(\hat p)\right) C^{--}(\hat q%
)\right]
\end{array}
\end{equation}
\begin{equation}
\label{sigm}
\begin{array}{c}
\eta ^{-}(
\hat k)=\frac 12k^2\int_{\hat p+\hat q=\hat k}d\hat p\times  \\ \left[
b_3(k,p,q)\left( G^{-+}(\hat p)+G^{--}(\hat p)\right) C^{++}(\hat q%
)+b_4(k,p,q)\left( G^{++}(\hat p)+G^{+-}(\hat p)\right) C^{+-}(\hat q%
)\right]
\end{array}
\end{equation}
Here $b_i(k,p,q)=k^{-2}B_i(k,p,q)$, and $B_i$'s are \cite{Leslie}%
$$
\begin{array}{ll}
B_1(k,p,q)= & kp(-z+z^3+x^2z+xyz^2) \\
B_2(k,p,q)= & kq(1+z^2)(z+xy) \\
B_3(k,p,q)= & kp(-z+z^3+xy+x^2z+y^2z+xyz^2) \\
B_4(k,p,q)= & kp(-z+z^2+y^2z+xyz^2)
\end{array}
$$
where $(x,y,z)$ are the cosines of angles between ${\bf (p,q),(q,k)}$, and $%
{\bf (k,p)}$ respectively.

When ${\bf z}^{+}{\bf =z}^{-},$ we have $C^{++}=C^{--}=C^{+-},G^{++}=G^{--},$
and $G^{+-}=G^{-+}$. Under this condition we expect $\Sigma ^{++}=\Sigma
^{--}$, but Eqs. (\ref{sigma},\ref{sigp},\ref{sigm}) do not yield this
results. However, the equality can be easily achieved by symmetrizing Eqs. (%
\ref{sigp},\ref{sigm}) by interchange of + and $-$ signs in Eq. (\ref{Gfull}%
) and averaging of the resulting equations with corresponding Eqs.
(\ref{sigp}) and (\ref{sigm}).  The final equations for $\eta^{\pm}$ are
\begin{equation}
\label{sig}
\begin{array}{c}
\eta ^{\pm }(
\hat k)=\lim_{\omega \rightarrow 0}\frac 12k^2\int d\hat p\times \\ \times
[b_1(k,p,q)\left( G^{\mp \pm }(
\hat p)+G^{\mp \mp }(\hat p)\right) C^{+-}(\hat q) \\ +b_2(k,p,q)\left(
G^{\pm \pm }(
\hat p)+G^{\pm \mp }(\hat p)\right) C^{\mp \mp }(\hat q) \\
+b_3(k,p,q)\left( G^{\pm \pm }(
\hat p)+G^{\pm \mp }(\hat p)\right) C^{\mp \mp }(\hat q) \\
+b_4(k,p,q)\left( G^{\mp \mp }(\hat p)+G^{\mp \pm }(\hat p)\right) C^{+-}(%
\hat q)]
\end{array}
\end{equation}

Now when ${\bf z}^{+}{\bf =z}^{-}$, we recover the fluid limit. In this case $%
C^{++}=C^{--}=C^{+-}$ and $\Sigma ^{++}=\Sigma ^{--}=$ $2\Sigma ^{fluid}$.
Here we take $\omega \longrightarrow 0$ limit in our calculation; this
approximation is valid for the large and inertial scale fluctuations. After
this we solve for $\Sigma $ and $G$ self consistently. A typical
self-energy digram is shown in Figure 2. Considering that $%
{\bf z}^{+}={\bf z}^{-}$ correspond to fluid limit, we postulate the
following relaxation behaviour for the $z-z$ correlation functions:
\begin{equation}
\label{corr}C^{\pm \pm }(k,\omega )=\frac 1{2\pi }\frac{C^{\pm }(k)}{%
(i\omega -\eta _k^{\pm })};C^{\pm \mp }(k,\omega )=\frac 1{2\pi }\frac{C^R(k)%
}{(i\omega -\frac{\eta _k^{+}+\eta _k^{-}}2)},
\end{equation}
\begin{equation}
\begin{array}{c}
C^{\pm \pm }(k,t_2-t_1)=C^{\pm }(k)\exp \left( -\eta _k^{\pm }\left(
t_2-t_1\right) \right) ; \\
C^{\pm \mp }(k,t_2-t_1)=C^R(k)\exp \left( -\frac{\eta _k^{+}+\eta _k^{-}}2%
\left( t_2-t_1\right) \right) ;
\end{array}
\end{equation}
The Eq. (\ref{Gfull}) yields
\begin{equation}
G(\hat k)=\frac 1{-\omega \left( \omega +i\frac{\eta ^{+}+\eta ^{-}}2\right)
}\left[
\begin{array}{cc}
-i\omega +\frac{\eta ^{+}}2 & \frac{\eta ^{-}}2 \\
\frac{\eta ^{-}}2 & -i\omega +\frac{\eta ^{+}}2
\end{array}
\right]
\end{equation}
which after substitution in Eq. (\ref{sig}) and integration of $\omega
^{\prime }$ integral \cite{rg} yields
\begin{equation}
\label{etafinal}
\begin{array}{c}
\eta ^{\pm }(k)=\int d
{\bf p}\frac 2{\eta _k^{+}+\eta _k^{-}}k^2 \\ \times [
\frac{b_1(k,p,q)C^R(q)\eta _p^{\mp}}{\eta _p^{+}+\eta _p^{-}+\eta _q^{+}+\eta
_q^{-}}+\frac{b_2(k,p,q)C^{\mp }(q)\eta _p^{\pm}}{\eta _p^{+}+\eta
_p^{-}+2\eta _q^{\mp }} \\ +\frac{b_3(k,p,q)C^{\mp }(q)\eta _p^{\pm}}{\eta
_p^{+}+\eta _p^{-}+2\eta _q^{\mp }}+\frac{b_4(k,p,q)C^R(q)\eta _p^{\mp}}{\eta
_p^{+}+\eta _p^{-}+\eta _q^{+}+\eta _q^{-}}]
\end{array}
\end{equation}
We have ignored the contributions from the pole at $\omega ^{\prime }=0$
because it corresponds to a static situation. The quantities $\eta ^{+}/2$
and $\eta ^{-}/2$ correspond to the inverse of the response time of the
eddies $z^{+}$ and $z^{-}$ respectively.

Our self-energy matrix is differs from that obtained by Camargo and
Tasso's using renormalization group calculations  \cite{mhdrg};
their $\hat \Sigma $ is symmetrical
with both the diagonal elements equal. This is due to different assumptions
of the models. In Camargo and Tasso's model cross helicity is zero which
forces both the diagonal elements to be equal. However, our model can
be applied to situations where cross helicity is not zero.
Also, we assume independent $%
f^{+}$ and $f^{-}$ forcing, whereas Camargo and Tasso assume independent $%
f^u $ and $f^b$ forcing. Renormalization group calculation for varying
normalized cross helicity is under progress.

We need to evaluate the integrals in the above equation In 3D the integral
\cite{Leslie} is
\begin{equation}
\int d{\bf p}=\int \int dpdq\frac{2\pi pq}k
\end{equation}
while in 2D it is
\begin{equation}
\int d{\bf p}=\int \int dpdq\frac 1{\sin ({\bf p,q)}}.
\end{equation}
The integral is to be performed in the hatched region of Figure 3.

We substitute the power-law expression of $C^{+}(k),C^{-}(k),$ and $C^R(k)$
from Eqs. (\ref{Kolmlike},\ref{Ek3d},\ref{Ek2d}). For $\eta ^{\pm }(k)$,
which is interpreted as the inverse of time-scale of $z^{\pm }$, we
substitute \cite{Leslie}
\begin{equation}
\label{eta}\ \eta ^{\pm }(k)=\Lambda ^{\pm }\left( \Pi ^{\mp }\right)
^{2/3}\left( \Pi ^{\pm }\right) ^{-1/3}k^{2/3},
\end{equation}
where $\Lambda ^{\pm }$ are constants. We also assume that the inertial
range spectra of the residual energy $E^R(k)$ is proportional to $k^{-5/3}$.
The EDQNM calculations of Pouquet et al. and Grappin et al. \cite{Grappin}
show that $E^R(k)$ is proportional to $k^{-2}$. Marsch and Tu \cite
{observation} analysed the solar wind data and found that the residual
energy spectra is between $k^{-3/2}$ and $k^{-5/3}$, but closer to $k^{-3/2}$%
. Hence, there is no consistent theory for the spectra of the residual
energy. In our model, for consistency, we assume that $E^R(k)\propto $ $%
k^{-5/3}$. With this power law the ratio between $E^R\left( k\right) $ and
the total energy $E\left( k\right) $ is constant; we denote this constant by
$\alpha .$ Similarly the ratio $E^{-}(k)/E^{+}(k)$ is a constant in our
model; we denote it by $\beta $.

Now the substitution of power laws for $\eta _k^{\pm }$ in Eq. (\ref
{etafinal}) yields \cite{Leslie}
\begin{equation}
\label{self}
\begin{array}{c}
\frac{\left( \Lambda ^{\pm }\right) ^2}{K^{\mp }}=\int \int d\varsigma
d\kappa \varsigma \kappa ^{-8/3}\times \\ \left[ \frac 1{1+\xi ^{\pm 1}}%
\left( \frac{b_2(1,\varsigma ,\kappa )+b_3(1,\varsigma ,\kappa )}{\left(
1+\xi ^{\pm 1}\right) \varsigma ^{2/3}+2\xi ^{\pm 1}\kappa ^{2/3}}\right) +%
\frac{\alpha (1+\beta ^{\mp 1})}{2\left( 1+\xi ^{\mp 1}\right) }\left( \frac{%
b_1(1,\varsigma ,\kappa )+b_4(1,\varsigma ,\kappa )}{\left( 1+\xi ^{\pm
1}\right) \left( \varsigma ^{2/3}+\kappa ^{2/3}\right) }\right) \right]
\end{array}
\label{selffinaleq}
\end{equation}
in three dimensions.  Here
 $\xi =\left( \Lambda ^{-}\Pi ^{+}\right) /\left( \Lambda ^{+}\Pi
^{-}\right) ,p=\zeta k$ and $q=\kappa k$.  In 2D the terms $vw^{-8/3}$
and $wv^{-8/3}$ are replaced by $w^{-8/3}$ and $%
v^{-8/3}$ respectively, and the whole integrand is divided by $\pi
(1-x^2)^{1/2}$ \cite{Leslie,Kraich:71}.

The integrals of Eq.\ (\ref{sigma}) suffer from the well known ``infrared
problem'' which comes from the strong dynamic coupling of fluctuations with
widely differing wavenumbers. Over the years, various techniques have been
developed to tackle the difficulty in the context of pure fluid turbulence.
Simplest among all these methods is introduction of cutoff, which is
discussed in Leslie\cite{Leslie}. Later on Renormalization group technique
\cite{rg},  the Lagrangian or semi-Lagrangian pictures \cite{Lagrange}, and
self-consistent screening \cite{mode} were used to make the integral
naturally finite and
cut off independent. The infrarred difficulties associated with Eq.~(\ref
{self}) can be similarly resolved. Knowing that the full theory is
constrained to be finite, we adopt the practical procedure of evaluating the
integral in Eq.~(\ref{sigma}) with a cut off $k_0=\lambda k$ and choosing $%
\lambda =1$ so that the pure fluid value of $\Lambda ^2/K$ are obtained
correctly when $\alpha =\beta =1$. Thereafter $\lambda $ is not varied.

In 2D when $E^u<E^b$, we choose $\lambda =1$, same as $\lambda $ of 3D, but
not 0.065 which yields $K=6.6$ in fluid limit. The choice of same $\lambda $
in 2D MHD was motivated by fact that in MHD turbulence, forward cascade of
energy and inverse cascade of cross helicity occur in both 2D and 3D (refer to
Ting et al. \cite{dynamo}). Note that in 2D fluid turbulence an inverse cascade
of energy
occurs, hence the behaviour of fluids turbulence in 2D and in 3D are
dramatically different. It appears that $\lambda =1$ is not applicable for
cases when $E^u>E^b$; therefore we have calculated the constants for
$E^{u} > E^{b}$.

To obtain the numerical value of $K^{\pm }$ and $\Lambda ^{\pm }$ we need
two additional equation involving $K^{\pm }$ and $\Lambda ^{\pm }$. The
additional equations we use are the equations of cascade rates of $z^{\pm }$%
; these equations are derived in the next subsection.

\subsection{ Computation of Cascade rates}

The cascade rates or the fluxes $\Pi ^{\pm ,R}$ have already been defined in
terms of $T^{\pm ,R}$ respectively (ref. Eqs. (\ref{fluxpm},\ref{fluxR})). We
evaluate $T$ $^{\pm ,R}$ to first order, and then, following Kraichnan \cite
{Kraich:59,Leslie} substitute $G^0,C^0$'s by $G$'s and $C$'s respectively
(refer to Leslie \cite{Leslie} for this procedure applied in fluid
turbulence). We will symbolically illustrate the evaluation of $T$ up to
first order. For this part of the calculation we work in $({\bf k},t)$ space
(refer to Leslie (\cite{Leslie}) for details). The quantity $T$ is
proportional to equal time triple correlations. To first order $T$ is
\begin{equation}
\begin{array}{c}
T(k,t,t)\propto \int_{
{\bf p+q=k}}d{\bf p}\left\langle z^{(0)}({\bf k,}t)z^{(0)}({\bf p,}t)z^{(1)}(%
{\bf q},t)\right\rangle \\ +\left\langle z^{(0)}({\bf k,}t)z^{(1)}({\bf p,}%
t)z^{(0)}({\bf q},t)\right\rangle +\left\langle z^{(1)}({\bf k,}t)z^{(0)}(%
{\bf p,}t)z^{(0)}({\bf q},t)\right\rangle .
\end{array}
\end{equation}
Note that $\left\langle z^{(0)}({\bf k,}t)z^{(0)}({\bf p,}t)z^{(0)}({\bf q}%
,t)\right\rangle $ is zero. Now using Eq. (\ref{z1}) we expand $z^{(1)}$ and
substitute in the above equation, which yields
\begin{equation}
\begin{array}{c}
T(k,t,t)\propto \int_{
{\bf p+q=k}}d{\bf p}\int_{{\bf r+s=q}}d{\bf r}\int_{-\infty }^tdt^{\prime
}\times \\ \left\langle z^{(0)}({\bf k,}t)z^{(0)}({\bf p,}t)G^{(0)}({\bf q}%
,t-t^{\prime })z^{(0)}({\bf r},t^{\prime })z^{(0)}({\bf s},t^{\prime
})\right\rangle +\cdots
\end{array}
\end{equation}
or,
\begin{equation}
\label{T1st}T(k,t,t)\propto \int_{{\bf p+q=k}}d{\bf p}\int_{-\infty
}^tdt^{\prime }G^{(0)}({\bf q},t-t^{\prime })C({\bf k,}t-t^{\prime })C({\bf p%
},t-t^{\prime })+\cdots
\end{equation}
The terms $G$'s and $C$'s appear as $G^{s_1s_2}$ and $C^{s_1s_2}$ in the
final expression. The full expression for $T^{\pm }$ is
\begin{equation}
\label{Tfull}
\begin{array}{c}
T^{\pm }(k,t,t)=4\pi k^4\int_{
{\bf p+q=k}}d{\bf p}\int_{-\infty }^tdt^{\prime }\times \\ \lbrack
-b_1(k,p,q)\left\{ G^{\mp \pm }(p,\Delta t)C^R(q,\Delta t)C^{\pm }(p,\Delta
t)+G^{\mp \mp }(p,\Delta t)C^R(k,\Delta t)C^{\pm }(q,\Delta t)\right\} \\
-b_3(k,p,q)\left\{ G^{\mp \pm }(p,\Delta t)C^R(k,\Delta t)C^{\pm }(q,\Delta
t)+G^{\mp \mp }(p,\Delta t)C^R(q,\Delta t)C^{\pm }(k,\Delta t)\right\} \\
-b_2(k,q,p)\left\{ G^{\pm \mp }(q,\Delta t)C^R(p,\Delta t)C^R(k,\Delta
t)+G^{\pm \pm }(q,\Delta t)C^{\mp }(p,\Delta t)C^{\pm }(k,\Delta t)\right\}
\\
-b_4(k,q,p)\left\{ G^{\pm \mp }(q,\Delta t)C^{\mp }(p,\Delta t)C^{\pm
}(k,\Delta t)+G^{\pm \pm }(q,\Delta t)C^R(p,\Delta t)C^R(k,\Delta t)\right\}
\\
+b_2(k,q,p)\left\{ G^{\pm \mp }(k,\Delta t)C^R(p,\Delta t)C^R(q,\Delta
t)+G^{\pm \pm }(k,\Delta t)C^{\mp }(p,\Delta t)C^{\pm }(q,\Delta t)\right\}
\\
+b_1(k,p,q)\left\{ G^{\pm \mp }(k,\Delta t)C^{\mp }(p,\Delta t)C^{\pm
}(q,\Delta t)+G^{\pm \pm }(k,\Delta t)C^R(p,\Delta t)C^R(q,\Delta t)\right\}
]
\end{array}
\end{equation}
where $\Delta t=t-t^{\prime }$. The expression for $T^R$ is similar to $%
T^{\pm }$. Substitution of $T$'s yield the following cascade rates:
\begin{equation}
\Pi ^{\pm ,R}(k)=\int_k^\infty dk^{\prime }\int \int dpdqS^{\pm
,R}(k^{\prime },p,q)
\end{equation}
where the $dpdq$ integral is over the hatched region of Figure 3. Note
however that the hatched region A  has $q>p$. If we
have $p>k$, then all three vectors ${\bf k,p,}$and ${\bf q}$ forming
triangle will have magnitudes greater than $k$; hence these triad will not
contribute to the flux. Therefore, in region A the flux contribution comes
from the following region of integration:
\begin{equation}
\Pi ^{\pm ,R}(k)=\int_k^\infty dk^{\prime }\int_0^kdp\int_{p*}^{p+k^{\prime
}}dqS^{\pm ,R}(k^{\prime },p,q)
\end{equation}
where
\begin{equation}
p^{*}=\max (p^{\prime },|p-k^{\prime }|).
\end{equation}
For region B similar arguments lead to
\begin{equation}
\Pi ^{\pm ,R}(k)=\int_k^\infty dk^{\prime }\int_0^kdq\int_{q*}^{q+k^{\prime
}}dpS^{\pm ,R}(k^{\prime },p,q)
\end{equation}
where $q^{*}=\max (q^{\prime },|q-k^{\prime }|).$

After substitution of $C^{\pm ,R}$ and $G$ in the Eq. (\ref{T1st}) and with
the change of variable \cite{Leslie}
\begin{equation}
k^{\prime }=\frac ku;p=\frac{vk}u;q=\frac{wk}u
\end{equation}
we obtain
\begin{equation}
\label{pipm}\frac{\Lambda ^{\pm }}{K^{+}K^{-}}=\left[ \int dv\ln \frac 1v%
\int_{v^{*}}^{1+v}dw+\int dw\ln \frac 1w\int_{w^{*}}^{1+w}dv\right] \Psi
^{\pm }(1,v,w)
\end{equation}
\begin{equation}
\label{piR}
\begin{array}{c}
\Pi ^R=
\frac 12\frac{\Pi ^{+}K^{+}K^{-}}{\Lambda ^{+}}\left[ \int dv\ln \frac 1v%
\int_{v^{*}}^{1+v}dw+\int dw\ln \frac 1w\int_{w^{*}}^{1+w}dv\right]  \\
\times \left( \Psi ^{R1}(1,v,w)+\Psi ^{R2}(1,v,w)\right)
\end{array}
\end{equation}
where in 3D
\begin{equation}
\label{psip}
\begin{array}{c}
\Psi ^{\pm }(1,v,w)=-
\frac{vw^{-8/3}\alpha (1+\beta ^{\mp 1})/2}{(1+\xi ^{\mp 1})}\left(
b_1(1,v,w)+b_3(1,v,w)\right) \times  \\ \left[
\frac 1{\left( 2+(1+\xi ^{\pm 1})\left( v^{2/3}+w^{2/3}\right) \right) }+%
\frac 1{\left( (1+\xi ^{\pm 1})\left( 1+v^{2/3}\right) +2w^{2/3}\right) }%
\right]  \\ -
\frac{v^{-8/3}w}{(1+\xi ^{\pm 1})}\left( b_2(1,w,v)+b_4(1,w,v)\right) \times
\\ \left[
\frac 1{\left( 2+2\xi ^{\pm 1}v^{2/3}+(1+\xi ^{\pm 1})w^{2/3}\right) }+\frac{%
\alpha ^2(1+\beta )^2/(4\beta )}{\left( (1+\xi ^{\pm 1})\left(
1+v^{2/3}+w^{2/3}\right) \right) }\right]  \\ +
\frac{\left( vw\right) ^{-8/3}}{(1+\xi ^{\pm 1})}\left(
b_1(1,v,w)+b_2(1,w,v)\right) \times  \\ \left[ \frac 1{\left( 1+\xi ^{\pm
1}+2\xi ^{\pm 1}v^{2/3}+2w^{2/3}\right) }+\frac{\alpha ^2(1+\beta
)^2/(4\beta )}{\left( (1+\xi ^{\pm 1})\left( 1+v^{2/3}+w^{2/3}\right)
\right) }\right]
\end{array}
\end{equation}
\begin{equation}
\begin{array}{c}
\Psi ^{R1}(1,v,w)=-
\frac{vw^{-8/3}}{(1+\xi ^{-1})}\left( b_1(1,v,w)+b_3(1,v,w)\right) \times
\\ \left[
\frac{\alpha ^2(1+\beta )^2/(4\beta )}{\left( (1+\xi )\left(
1+v^{2/3}+w^{2/3}\right) \right) }+\frac 1{\left( 2\xi +(1+\xi
)v^{2/3}+2w^{2/3}\right) }\right]  \\ -
\frac{v^{-8/3}w\alpha (1+\beta )/2}{(1+\xi )}\left(
b_2(1,w,v)+b_4(1,w,v)\right) \times  \\ \left[
\frac 1{\left( 2\xi +2\xi v^{2/3}+(1+\xi )w^{2/3}\right) }+\frac 1{\left(
(1+\xi )\left( 1+w^{2/3}\right) +2\xi v^{2/3}\right) }\right]  \\ +
\frac{\left( vw\right) ^{-8/3}}{(1+\xi ^{-1})}\left(
b_1(1,v,w)+b_2(1,w,v)\right) \times  \\ \left[ \frac 1{\left( 1+\xi +2\xi
v^{2/3}+2w^{2/3}\right) }+\frac{\alpha ^2(1+\beta )^2/(4\beta )}{\left(
(1+\xi )\left( 1+v^{2/3}+w^{2/3}\right) \right) }\right]
\end{array}
\end{equation}
\begin{equation}
\begin{array}{c}
\Psi ^{R2}(1,v,w)=-
\frac{vw^{-8/3}}{(1+\xi )}\left( b_1(1,v,w)+b_3(1,v,w)\right)  \\ \times
\left[
\frac{\alpha ^2(1+\beta )^2/(4\beta )}{\left( (1+\xi )\left(
1+v^{2/3}+w^{2/3}\right) \right) }+\frac 1{\left( 2+(1+\xi )v^{2/3}+2\xi
w^{2/3}\right) }\right]  \\ -
\frac{v^{-8/3}w\alpha (1+\beta )}{(1+\xi ^{-1})2\beta }\left(
b_2(1,w,v)+b_4(1,w,v)\right)  \\ \times \left[
\frac 1{\left( 2+(1+\xi )\left( v^{2/3}+w^{2/3}\right) \right) }+\frac 1{%
\left( (1+\xi )\left( 1+w^{2/3}\right) +2\xi v^{2/3}\right) }\right]  \\ +
\frac{\left( vw\right) ^{-8/3}}{(1+\xi )}\left( b_1(1,v,w)+b_2(1,w,v)\right)
\\ \times \left[ \frac 1{\left( 1+\xi +2v^{2/3}+2\xi w^{2/3}\right) }+\frac{%
\alpha ^2(1+\beta )^2/(4\beta )}{\left( (1+\xi )\left(
1+v^{2/3}+w^{2/3}\right) \right) }\right] .
\end{array}
\end{equation}
In 2D the terms $vw^{-8/3}$ and $wv^{-8/3}$ are replaced by $w^{-8/3}$ and $%
v^{-8/3}$ respectively, and the whole integrand is divided by $\pi
(1-x^2)^{1/2}$ \cite{Leslie,Kraich:71}.

\subsection{Computation of $K^{\pm }$ and $\Lambda ^{\pm }$}

We numerically solve for $K^{\pm }$ and $\Lambda ^{\pm }$ from Eqs. (
\ref{selffinaleq}, \ref
{pipm}) for given ratios of $\alpha =E^R(k)/E(k)$ and $\beta
=E^{-}(k)/E^{+}(k)$. Since the integrand itself is a function of $\xi $,
which it self is a function of $\Lambda ^{\pm },\Pi ^{+},$ and $%
\Pi^{-}$, we start with a guessed value of $\xi $ and iterate until
the two successive values of $\xi $ are approximately equal. We perform
these calculations for both 2D and 3D. The values of $K^{\pm }$, $\Lambda
^{\pm },$ and $\xi $ for various $\alpha $ and $\beta $ for 3D are listed in
Table 1, whereas for 2D, they are listed in Table 2.

\subsection{Computation of $K^R$}

If we postulate that
\begin{equation}
E^R(k)=K^{R1}\left( \Pi ^R\right) ^{2/3}k^{-5/3},
\end{equation}
then the substitution of $\Pi ^R$ in Eq. (\ref{piR}) yields
\begin{equation}
K^{R1}=K^{+}\left( \frac{E^R(k)}{E^{+}(k)}\right) \left( \frac{\Lambda ^{-}}{%
K^{+}K^{-}I}\right) ^{2/3}
\end{equation}
where $I$ is the value of the integral of Eq. (\ref{piR}). The values of $%
K^{R1} $ computed for various $\alpha $ and $\beta $ are listed in Tables 1 and
2.

We can model $E^R(k)$ in yet another way. This modeling is motivated by Eq. (%
\ref{corrR}). From this equation we argue that
\begin{equation}
\Pi ^R\sim \dot E^R\sim k\left( \frac{z_k^{+}+z_k^{-}}2\right) \left(
z_k^{+}z_k^{-}\right) .
\end{equation}
Therefore,
\begin{equation}
\Pi ^R=k\left( \frac{z_k^{+}+z_k^{-}}2\right) \frac{E^R(k)k}{K^{R2}}
\end{equation}
or,
\begin{equation}
E^R(k)=\frac{2K^{R2}k^{-5/3}\Pi ^R}{\sqrt{K^{+}}\left( \Pi ^{-}\right)
^{2/3}\left( \Pi ^{+}\right) ^{-1/3}+\sqrt{K^{-}}\left( \Pi ^{+}\right)
^{2/3}\left( \Pi ^{-}\right) ^{-1/3}}.
\end{equation}
The constants $K^{R2}$ calculated using the above equation are listed in Tables
1 and 2.

\subsection{Discussion}

Several important conclusions can be drawn from the numbers shown in Tables
1 and 2. The Kolmogorov's constant for MHD turbulence $K^{\pm }$ are not
universal constants as compared to universal fluid Kolmogorov's constant. We
find that the constants $K^{\pm }$ depend on Alfv\'en ratio $r_A$ and ratio
of $E^{-}(k)$ and $E^{+}(k)$. They could also depend on the mean magnetic
field, but study of this effect is beyond the scope of this paper.

As reported in our earlier paper \cite{Verma:dia}, when $\beta
=E^{-}(k)/E^{+}(k)=1$, the constants $K^{+}=K^{-}=K$ increases monotonically
from 1.43 to $\sim $4.07 as we go from fully fluid case $r_A=\infty $ to $%
r_A\sim 0.2$, then it decreases and finally reaches 3.51 when the plasma is
fully magnetic. Please note that the numbers here are slightly different
from those in \cite{Verma:dia}; the discrepancy is due to the lower
accuracy used in the earlier paper. In 2D when $\beta =1$, $K$ is in the
range of $5-7.5$ for $0<r_A<1$. In both 2D and 3D for a given $r_A$, as we
decrease $\beta $ (or increase normalized cross helicity $\sigma _c$), $K^{+}
$ (constant corresponding to larger of $E^{+}$ and $E^{-}$)
decreases whereas $K^{-}$ increases.

The above DIA results are in qualitative agreement with the preliminary
numerical results of Verma et al. \cite{Verma:sim}. Verma et al. performed
direct numerical simulation of MHD turbulence in 2D (8 runs) and 3D (single
run) for various initial normalized cross helicity and mean magnetic field.
When $E^{+} \approx E^{-}$, the constant $K^{+}$ was approximately
equal to $K^{-}$.
The mean values of the constant obtained by them were $%
K=3.7$ in three-dimensions  for $E^{-}/E^{+}=0.6$, and $K=6.6$
in two-dimensions for $E^{-}/E^{+}=1,0.6$.
However, the constant $K$ of the majority species (larger of $z^{\pm }$) was
always marginally lower
than that of minority species. Unfortunately, we do not have the inertial
range Alfv\'{e}n ratio $r_{A}(k)$ for our simulations; we will study the
$r_{A}$ dependence in future simulations.  However, we supsect
the inertial range Alfv\'en ratio to be
in the range of $0.25-1$. We find that
for small $\sigma_{c}$ the mean values of
$K$ from the simulations are generally close to the  DIA
result for $r_{A}=1$ and $\sigma_{c}=0$.

However, for large $\sigma _c$ there appears to be serious
discrepancies between the simulation results and our results.
In simulations for large $\sigma_c$, $\Pi ^{+}/\Pi ^{-}\sim
E^{-}(k)/E^{+}(k)$, however, in our DIA calculations $\Pi ^{+}/\Pi ^{-}\ll
E^{+}(k)/E^{-}(k)$ in fluid dominated cases and $\Pi ^{+}/\Pi ^{-}\gg
E^{+}(k)/E^{-}(k)$ in equipartioned ($r_A=1$) or magnetically dominated
case.   Regarding constants $K^{\pm}$, for $\sigma_{c}(k)
\approx 0.9 (\beta=0.05)$ Verma et al.\ found that the average $K^{+} \approx
2.4$ and $K^{-}
\approx 24.0$.  In DIA calculations, for magnetically dominated cases, the
equations for $K^{\pm}$ do not yield real solutions.
 This indicates that our model is not fully consistent atleast
in magnetically dominated limit; this point is illustrated in the
following paragraph. Nevertheless, there are some qualititative agreement
between the simulation results and DIA results, e.g., the constant $K$
corresponding
to the majority species is always greater than the $K$ corresponding
to the minority species.

When we turn off the velocity field, i. e.,  ${\bf z}^{+}=-{\bf z}^{-}={\bf b%
}$, the MHD equations are linear. Here the magnetic energy dissipates only due
to the resistivity, and there is no turbulent cascade rate. However, in our
formalism, we obtain nonzero turbulent cascade rate for this case. Also, there
is no
consistent solution for $K^{\pm }$ when $\sigma _c$ is large and $r_A\leq 1$%
. To circumvent this inconsistencies we may have to modify our relaxation
time, Green's functions etc., and also possibly may need to use $%
{\bf u}$ and ${\bf b}$ rather ${\bf z}^{\pm }$. This work is under progress.

Note that the ratio $\Pi ^R/\Pi ^{+}$ increases monotonically as $r_A$
decreases from infinity to zero. By the time $r_A=5$, the
ratio $\Pi ^R/\Pi ^{+}$ has already become large. This large ratio implies
that $\Pi ^b\approx -\Pi ^u$ ($\Pi ^u>0$ and $\Pi ^b<0$), i.e., the kinetic
energy is cascading from small $k$ to large $k$, while the magnetic energy
is cascading from large $k$ to small $k$. This is a theoretical
demonstration of inverse cascade of magnetic energy.  Earlier work on
amplification of magnetic energy have been done by Fyfe et al.\ using
numerical simulations,
L\'{e}orat et al. using EDQNM closure scheme, Gloguen
et al.\  using scalar model of MHD turbulence, and
Ting et al.\ by turbulent relaxation (\cite{dynamo} and references therein).
Our results may be useful for modeling
dynamos from turbulent energy cascade point of view. If we start
with small magnetic field in a fluid dominated plasma, there will be a small
flux of magnetic energy cascading to large-scales from small-scales; that
will enhance the magnetic field in the system and decrease $r_A$. As $r_A$
decreases, the flux cascading inversely also increases, hence further
enhancing the magnetic field of the system. After the magnetic field has
become sufficiently large ($r_A=5$), almost all the fluid energy flux goes
into magnetic energy flux, which makes the enhancement of the magnetic field
faster. A quantitative model based on these ideas could be useful in
estimating the time-scales etc.\ in dynamos.

In the above discussion we have not differentiated between inertial range $%
r_A$ ($E^u(k)/E^b(k)$) and the $r_A=E^u/E^b$, which may be a gross
approximation. Also modeling of relaxation of $C^{+-}(k,\Delta t)$ as well
as modeling of $E^R(k)$ is not on a strong footing. We do not get consistent
answers for equipartioned and magnetically dominated cases. Numerical
simulations might provide important clues that could help us modeling these
quantities better.

In the next section we will apply DIA to MHD turbulence in situations when $%
C_A\gg z^{\pm }$.

\pagebreak

\section{DIA OF MHD with $C_A\gg z^{\pm }$}

As mentioned in the introduction, in presence of a strong mean magnetic field,
the energy spectra is expected to be proportional to $k^{-3/2}$. However,
the solar wind observations \cite{observation} as well as the numerical
simulations \cite{Verma:sim} do not appear to support this hypothesis at
least for $C_A/z^{\pm }\leq 5$. It is possible that when $C_A/z^{\pm }$ $\gg
1$, e.g., when the ratio is of the order of 100, the Kraichnan's \cite
{Kraich:mhd} or Dobrowolny et al.'s \cite{Dobro} arguments are valid.
Unfortunately numerical simulations have not yet probed in these regimes.

There exist another possibility that $B_0$ or $C_A$ which should be
substituted in Eq. (\ref{Kraich}) is not the mean magnetic field but the
renormalized $C_A$. When the mean magnetic field term of Eq. (\ref{mhdeqn0})
is comparable to the nonlinear term, simple dimensional arguments yields the
following scaling for $C_A$:
\begin{equation}
\label{renCA}C_A=\Pi ^{1/3}k^{-1/3}.
\end{equation}
where $\Pi $ is the total energy cascade rate. This argument is similar to
that due to Bhattacharjee \cite{jkb:sound} for renormalization of sound speed
in randomly stirred fluids. We conjecture that the small-scale fluctuations
are affected by renomalized magnetic field rather than the mean magnetic
field. Hence, we should substitute renormalized $C_A$ of Eq. (\ref{renCA})
in Eq. (\ref{Kraich}) that immediately yields $k^{-5/3}$ energy spectra for
energy. Therefore, large $C_A$ could yield $k^{-5/3}$ energy spectra
rather than $k^{-3/2}$ as argued by Kraichnan \cite{Kraich:mhd} and
Dobrowolny et al. \cite{Dobro}. However, in absence of a quantitative theory
or definite numerical results, we will make certain assumptions and calculate
the
Kraichnan's constant for a variation of Dobrowolny et al.'s model.

As argued by Kraichnan \cite{Kraich:mhd} and Dobrowolny et al. \cite{Dobro}
we assume that the Alfv\'en time-scale, rather that the nonlinear
time-scale, is the relaxation time-scale when $C_A/z^{\pm } \gg 1$. Using
random phase approximation Veltri et al. \cite{Veltri} showed that Green's
function and relaxation of correlation functions are
\begin{equation}
\hat{G}(k,t_2-t_1)=\left[
\begin{array}{cc}
\exp \left[ -gkC_A(t_2-t_1)\right]  & 0 \\
0 & \exp \left[ -gkC_A(t_2-t_1)\right]
\end{array}
\right]
\end{equation}
\begin{equation}
C^{\pm ,R}(k,t_2-t_1)=C^{\pm ,R}(k,t_1=t_2)\exp \left[
-gkC_A(t_2-t_1)\right] .
\end{equation}
We assume $g=1$. Unfortunately, Veltri et al. \cite{Veltri} assumed
equipartition of magnetic and fluid energy which is not the case in our
model discussed in this section. In Dobrowolny et al.'s model \cite{Dobro}
the dissipation rates $\Pi ^{+}$ and $\Pi ^{-}$ are equal irrespective of $%
E^{-}(k)/E^{+}(k)$ ratios. This has been found to be inconsistent with the
numerical results \cite{Verma:sim}. To bring in some effects of $%
E^{-}(k)/E^{+}(k)$ we modify the Dobrowolny et al.'s model in the following
manner:
\begin{equation}
\label{modifiedD}E^{\pm }(k)=A^{\pm }\left( \Pi ^{\pm }C_A\right)
^{1/2}k^{-3/2}
\end{equation}

We do not need to solve for self-energy in this case; Green's function has
been obtained using random phase approximation. We, however, solve for
cascade rates. The equation corresponding to Eq. (\ref{Tfull}) is
\begin{equation}
\label{TfullKr}
\begin{array}{c}
T^{\pm }(k,t,t)=4\pi k^4\int_{
{\bf p+q=k}}d{\bf p}\int_{-\infty }^tdt^{\prime }\times \\ \lbrack
-b_1(k,p,q)G(p,\Delta t)C^R(k,\Delta t)C^{\pm }(q,\Delta t) \\
-b_3(k,p,q)G(p,\Delta t)C^R(q,\Delta t)C^{\pm }(k,\Delta t) \\
-b_2(k,q,p)G(q,\Delta t)C^{\mp }(p,\Delta t)C^{\pm }(k,\Delta t) \\
-b_4(k,q,p)G(q,\Delta t)C^R(p,\Delta t)C^R(k,\Delta t) \\
+b_2(k,q,p)G(k,\Delta t)C^{\mp }(p,\Delta t)C^{\pm }(q,\Delta t) \\
+b_1(k,p,q)G(k,\Delta t)C^R(p,\Delta t)C^R(q,\Delta t)]
\end{array}
\end{equation}
where $G(k,\Delta t)=\exp (-C_At)$. From the Eqs. (\ref{TfullKr}) and a
similar equation for $T^R$ we can derive expressions $\Pi ^{\pm }$ and $\Pi
^R$.

{}From the equations for $\Pi ^{\pm }$ and $\Pi ^R$ we can solve for $A^{\pm }$
using manipulations similar to those used in the previous section. The
values of $A^{\pm }$ for various values of $\alpha $ and $\beta $ are listed
in Tables 1 and 2. When $\beta =1$, the constants $A^{+}=A^{-}=A$ are
approximately 2.0-2.5 in 3D and 3.5-4.0 in 2D. The constant $A$ obtained
here for 3D is close to the one obtained by Matthaeus and Zhou \cite
{Kolm-like}. As we vary $\beta $, the constants $A^{+}$ and $A^{-}$ begin to
differ. We find that when $E^{-}(k)<E^{+}(k),$ $A^{+}$ is always larger than
$A^{-}$ contrary to the Kolmogorov-like models in which $K^{+}<K^{-}$. For $%
r_A<1$, the ratio $\Pi ^{+}/\Pi ^{-}$ is less than one, whereas for $r_A>1$,
the ratio is greater than one. Hence $r_A$ appears to have significant
effect on cascade rates in this model as well.

The ratio $\Pi ^R/\Pi ^{+}$ decreases from a very large value (2519) to
nearly zero as $r_A$ varies from zero to one. This implies that there is an
inverse cascade of magnetic energy for $r_A<1$. As $r_A$ increases from 1,
the ratio $\Pi ^R/\Pi ^{+}$ decreases from zero to negative values (the
magnitude of the ratio increases) till $r_A$ reaches around 15, then it
again starts increasing. It reaches nearly zero at $r_A=1000$ and reaches
one at $r_A=\infty$.  This is an indication of inverse cascade of kinetic
energy for $1<r_A<1000$.  Qualitatively, when the magnetic energy dominates
the plasma, there is an inverse cascade of magnetic energy.  On the contrary,
when the kinetic energy dominates (only till $r_{A}=1000$), there is an
inverse cascade of kinetic energy.
These predictions differ from those in section 3. Recall that
in Kolmogorov-like models of the previous section, for all $r_A$ (except at $%
\infty $) we had an inverse cascade of magnetic energy.

Our generalized Kraichnan model is heuristic. Numerical simulations could
help us in determining which of the MHD turbulence models are applicable in
various parameter space. Comparisons of the numerical results with the above
predictions will provide insights into these puzzles.

\section{CONCLUSIONS}

In this paper we have applied direct-interaction approximation to MHD
turbulence
and obtained the values of the Kolmogorov's constants for MHD and Kraichnan's
constant for various $E^{-}(k)/E^{+}(k)$ and $E^u(k)/E^b(k)$ ratios. Ours is
a field-theoretical (nonvanishing) first order calculation. We have obtained
equations for the constants as well as ratios of energy cascade rates for
both Kolmogorov's and Kraichnan's power laws. We have solved for these
constants for various values of $E^{-}(k)/E^{+}(k)$ and $E^u(k)/E^b(k)$
ratios. When $\sigma _c=0,$ in three-dimensions the constants $K^{+}=K^{-}=K$
are close to 2 for fluid dominated cases, but between 3 and 4 for  cases
when $r_A\leq 1$. In 2D, however, $K$ is in the range of $5-7.5$ for $0<r_A<1$%
. For a given $r_A$ as we increase $\sigma _c$ (or decrease $\beta $), $K^{+}
$ (the constant corresponding to majority species) decreases whereas $K^{-}$
increases.

The DIA results for Kolmogorov-like phenomenology with small $\sigma _c$ are
in good agreement with the simulation results of Verma et al.\ \cite
{Verma:sim}. However, the DIA results for large $\sigma _c$ are only
qualitatively consistent with the numerical results. For example, the
Kolmogorov's constant for ($K^{+}$ here) the majority species is always
smaller than that for minority species, consistent with the numerical
results. However, for large $\sigma _c$ our DIA calculations yield $\Pi
^{+}/\Pi ^{-}\ll E^{+}(k)/E^{-}(k)$ in fluid dominated cases and $\Pi
^{+}/\Pi ^{-}\gg E^{+}(k)/E^{-}(k)$ in magnetically dominated cases; these
results
are in disagreement with the simulation results where $\Pi ^{+}/\Pi
^{-}\sim E^{+}(k)/E^{-}(k)$.  We believe that the inconsistency can
probably be removed if we modify the relaxation time or $E^{R}$ expressions,
or possibly we may have to use $u$ and $b$ varibles.
However,
note that in 3D simulation only a single run with a relatively lower
resolution ($128^3$) was performed. We need to perform more runs to compare
the DIA results with the simulations.

In this paper for the first time we have calculated Kraichnan's constants
for a generalization of Dobrowolny et al.'s model. We find that when $\beta
=1$, the constant $A^{+}=A^{-}=A$ is approximately 2.0-2.5 in 3D and 3.5-4.0
in 2D, consistent with earlier estimates of $A$. As $\beta $ decreases, the
constants $A^{+}$ and $A^{-}$ begin to differ. The DIA results for
Kraichnan's model  are inconsistent with the results of the numerical
simulations performed so far. It is possible that Kraichnan's model or its
generalizations becomes applicable for large $C_A/z^{\pm }$, however, at
present we do not know the critical $C_A/z^{\pm }$ where transitions from
Kolmogorov-like models to Kraichnan's model take place. Another possibility
is that $C_A$ of Kraichnan's model is the renormalized mean magnetic field
which is scale dependent ($\propto k^{-1/3}$), and that will yield $k^{-5/3}$
energy spectra.

Our DIA calculations using Kolmogorov-like energy spectra exhibits
an inverse cascade of magnetic energy flux.  We also find that as
$r_{A}$ decreases, the inverse cascade rate of magnetic energy increases.
This is a theoretical demonstration of inverse cascade of magnetic energy,
and this result could  be useful for modeling
dynamos from turbulent energy cascade point of view. However, DIA calculations
using the generalized Dobrowolny et al.'s model yields
an inverse cascade of magnetic energy for $r_A<1$, and
an inverse cascade of kinetic energy for $1<r_A<1000$.

The results discussed in this paper is useful for various applications in
the solar wind. For example, with the knowledge of the Kolmogorov's
constants for MHD, Verma et al. \cite{Verma:temp} have calculated the
contribution of turbulent dissipation to the overall heating in the solar
wind. We anticipate that similar theoretical calculations can be useful in
understanding of inverse cascade of energy, process called dynamo mechanism,
and inverse cascade of cross helicity, a process called dynamic alignment.

\pagebreak

\newpage

\begin{figure}
\figure{Figure 1: Diagramatic representation of the nonlinear interaction terms
 $s^{+}(.,.,.)={\bf (k \cdot z^{+}(a)) (z^{-}(b) \cdot \nabla z^{+}(c))}$,
where ${\bf a,b,c=(k,p,q)}$ (refer to Eq. (\ref{detail})).
The solid lines represent $z^{+}$ and the wavy lines represent $z^{-}$.}
\end{figure}

\begin{figure}
\figure{Figure 2: Diagramatic representation of self-energy $\Sigma(k,\omega)$.
The quantities $G$ and $C$ in the figure are dressed quantities.}
\end{figure}

\begin{figure}
\figure{Figure 3: The region of integration $\int_{{\bf p=q=k}} d{\bf p}$. }
\end{figure}

\begin{table}
\begin{center}
\begin{tabular}{||c|c||c|c|c|c|c|c|c|c|c||c|c|c|c||} \hline
\multicolumn{2}{||c||}{} &
\multicolumn{9}{c||}{Kolmogorov-like phenomenology} &
\multicolumn{4}{c||}{Kraichnan} \\ \hline
$r_{A}$ & $E^{-}/E^{+}$ & $\zeta$ & $\Lambda^{+}$ & $\Lambda^{-}$ &
$K^{+}$ & $K^{-}$ & $\Pi^{+}/\Pi^{-}$  & $K^{R1}$ & $K^{R2}$ &
$\Pi^{R}/\Pi^{+}$ &
$A^{+}$ & $A^{-}$ & $\Pi^{+}/\Pi^{-}$ & $\Pi^{R}/\Pi^{+}$ \\   \hline \hline
$\infty$ & 1 & 1& 0.390 & 0.390 & 1.43 & 1.43 & 1 & 1.43 & 1.20 & 1 &
 2.51 & 2.51 & 1 & 1 \\ \hline
1000 & 1 & 1 & 0.390 & 0.390 & 1.44 & 1.44 & 1 & 1.43 & 1.19 & 1 &
  2.51 & 2.51 & 1 & 0.063 \\ \hline
100 & 1 & 1 & 0.395 & 0.395 & 1.46 & 1.46 & 1 & 1.30 & 1.02 & 1.14 &
  2.48 & 2.48 & 1 & $-7.88$ \\ \hline
15 & 1 & 1 & 0.425 & 0.425 & 1.60 & 1.60 & 1 & 0.359 & 0.144 & 7.68 &
  2.31 & 2.31 & 1 & $-42.92$ \\ \hline
  & 1 & 1 & 0.491 & 0.491 & 1.92 & 1.92 & 1 & 0.0831 & 0.015 & 60.3  &
   2.12 & 2.12 & 1 & $-74.0$  \\ \cline{2-15}
5 & 0.5 & 0.51 & 0.775 & 0.254 &  1.69 & 2.05 & 1.55 & 4.15 & 0.96 & 92.1 &
   2.95 & 1.56 & 1.12 & $-80.0$\\ \cline{2-15}
   & 0.1 & 0.21 & 0.856 & 0.054 & 1.20 & 1.34 & 3.31 & 1.42 & 0.526 & 171 &
    6.43 & 1.14 & 3.13 & $-203$ \\ \hline \hline
  & 1   & 1      & 0.612 & 0.612 & 2.58 & 2.58 & 1 & 0.018 & 0.016 & 345 &
     1.98 & 1.98 & 1.0 & $-65.3 $ \\ \cline{2-15}
2 & 0.5 & 0.39 & 1.20 & 0.243 & 1.92 & 3.47 & 1.93 & 0.0099 & 0.018 & 341 &
     2.78 & 1.42 & 1.05 & $-71.6$ \\ \cline{2-15}
   & 0.1 & 0.12 & 1.57 & 0.045 & 1.58 & 2.96 & 4.22 & 0.0055 & 0.020 & 386 &
     5.82 & 0.685 & 1.39 & $-167$\\ \hline \hline
   & 1 &1   & 0.748 & 0.748 & 3.38 & 3.38 & 1 & 0 & 0 & 1092 &
         1.99 & 1.99 & 1.0 & $-0.773$  \\ \cline{2-15}
1 & 0.5 & 0.19 & 3.55 & 0.112 & 0.995 & 17.7 & 6.04 & 0 & 0 & 796 &
       2.82 & 1.41 & 1.0 & $-0.773$ \\ \cline{2-15}
   & 0.4 & 0.08 & 7.36 & 0.034 & 0.536 & 53.8 & 16.0 & 0 & 0 & 813 &
       3.15 & 1.26 & 1.0 & $-0.773$ \\ \cline{2-15}
   & 0.1 & - & - & - & - & - & - & - & - & - &
       6.30 & 0.630 & 1.0 & $-0.773$ \\ \hline \hline
     & 1    & 1     & 0.865 & 0.865 & 4.04 & 4.04 & 1      & -0.0075 & -0.0003
& 2391 &
        2.13 & 2.13 & 1 & 148.3 \\ \cline{2-15}
0.5 & 0.9 & 0.66 & 1.62  & 0.475 & 2.00 &  9.11 & 2.25 & -0.0044 & -0.0007 &
1749 &
        2.25 & 2.02 & 0.992 & 149.4 \\ \cline{2-15}
     & 0.5 & - & - & - & - & - & - & - & - & - &
        3.09 & 1.50 & 0.944 & 175.2  \\ \cline{2-15}
     & 0.1 & - & - & - & - & - & - & - & - & - &
        9.43 & 0.717 & 0.577 & 880.4 \\ \hline \hline
0.2 & 1 & 1 & 0.913 & 0.913 & 4.07 & 4.07 & 1 & -0.012 & -0.0004 & 3588 &
       2.54 & 2.54 & 1 & 527.4 \\ \hline
0    & 1 & 1 & 0.889 & 0.889 & 3.51 & 3.51 & 1 & -0.014 & -0.0005 & 3965 &
        4.13 & 4.13 & 1.0 & 2519 \\ \hline \hline
\end{tabular}
\end{center}
\caption{For 3D MHD turbulence Kolmogorov's constants $K^{\pm}, \Lambda^{\pm},
A^{\pm}$ etc.\  for various values of $r_{A}$ and $E^{-}/E^{+}$.}
\end{table}

\newpage\

\begin{table}
\begin{center}
\begin{tabular}{||c|c||c|c|c|c|c|c|c|c|c||c|c|c|c||} \hline
\multicolumn{2}{||c||}{} &
\multicolumn{9}{c||}{Kolmogorov-like phenomenology} &
\multicolumn{4}{c||}{Kraichnan} \\ \hline
$r_{A}$ & $E^{-}/E^{+}$ & $\zeta$ & $\Lambda^{+}$ & $\Lambda^{-}$ &
$K^{+}$ & $K^{-}$ & $\Pi^{+}/\Pi^{-}$  & $K^{R1}$& $K^{R2}$ & $\Pi^{R}/\Pi^{+}$
&
$A^{+}$ &$ A^{-} $& $\Pi^{+}/\Pi^{-}$ & $\Pi^{R}/\Pi^{+}$ \\   \hline \hline
5  & \multicolumn{10}{c||}{$\lambda = 1.0$ not applicable} &
3.52 & 3.52 & 1.0 & $-76.3$  \\ \hline \hline
   & 1 & 1 & 0.590 & 0.590 & 5.27 & 5.27 & 1 & 0 & 0 & 1256 &
     3.36 & 3.36 & 1 & $-0.757$ \\ \cline{2-14}
1 & 0.5 & 0.205 & 2.79 & 0.092 & 1.47 & 28.2 & 6.24 & 0 & 0 & 864 &
      4.74 & 2.37 & 1 & $-0.757$ \\  \cline{2-14}
   & 0.1 & - & - & - & - & - & - & - & - & - &
      10.6 & 1.06 & 1 & $-0.757$  \\ \hline \hline
     & 1    & 1     & 0.717 & 0.717 & 7.11 & 7.11 & 1 & -0.011 & -0.0003 & 3348
&
        3.66 & 3.66 & 1.0 & 163 \\ \cline{2-14}
0.5 & 0.8 & 0.46 & 8.70 & 0.112 & 0.467 & 475 & 35.7 & -0.0010 & 0.067 & 1757 &
        5.22 & 4.02 & 0.926 & 677 \\ \cline{2-14}%
     & 0.1 & - & - & - & - & - & - & - & - & - &
        19.7 & 1.23 & 0.392 & 1440 \\ \hline \hline
0    & 1 & 1 & 0.796 & 0.796 & 7.49 & 7.49 & 1 & -0.0195 & -0.0004 & 7532 &
        10.16 & 10.16 & 1 & 5714\\ \hline \hline
\end{tabular}
\end{center}
\caption{For 2D MHD turbulence Kolmogorov's
constants $K^{\pm}, \Lambda^{\pm},
A^{\pm}$ etc.\  for various values of $r_{A}$ and $E^{-}/E^{+}$.  The constant
$\lambda$ which appears in the
lower limit of the self-energy integral is chosen to be
1.0. }
\end{table}

\end{document}